\documentclass[article,twocolumn,nofootinbib,superscriptaddress]{revtex4-2}
\usepackage[colorlinks=true,linkcolor=blue]{hyperref}%
\usepackage{physics}
\usepackage{graphicx}
\usepackage{xcolor}
\usepackage[normalem]{ulem}

\begin{document}
		
\title{Self-consistent microscopic calculations for electron captures on nuclei in core-collapse supernovae}

\author{A. Ravli\'c}
\email{ravlic@frib.msu.edu}
\affiliation{Facility for Rare Isotope Beams, Michigan State University, East Lansing, Michigan 48824, USA} 

\author{S. Giraud}
\affiliation{Facility for Rare Isotope Beams, Michigan State University, East Lansing, Michigan 48824, USA}

\author{N. Paar}
\affiliation{Department of Physics, Faculty of Science, University of Zagreb, Bijeni\v cka c. 32, 10000 Zagreb, Croatia}

\author{R. G. T. Zegers }
\email{zegers@frib.msu.edu}
\affiliation{Facility for Rare Isotope Beams, Michigan State University, East Lansing, Michigan 48824, USA}
\affiliation{Department of Physics and Astronomy, Michigan State University, East Lansing, Michigan 48824, USA}

\date{\today} 
	
\begin{abstract}
Calculations for electron capture rates on nuclei with atomic numbers between $Z=20$ and $Z=52$ are performed in a self-consistent finite-temperature covariant energy density functional theory within the relativistic quasiparticle random-phase approximation. Electron captures on these nuclei contribute most to reducing the electron fraction during the collapse phase of core-collapse supernovae. 
The rates include contributions from allowed (Gamow-Teller) and first-forbidden (FF) transitions, and it is shown that the latter become dominant at high stellar densities and temperatures. Temperature-dependent effects such as Pauli unblocking and transitions from thermally excited states are also included. The new rates are implemented in a spherically symmetric 1D simulation of the core-collapse phase. The results indicate that the increase in electron capture rates, due to inclusion of FF transitions, leads to reductions of the electron fraction at nuclear saturation density, the peak neutrino luminosity, and enclosed mass at core bounce. The new rates reaffirm that the most relevant nuclei for the deleptonization situate around the $N = 50$ and $82$ shell closures, but compared to previous simulations, nuclei are less proton rich. The new rates developed in this work are available, and will be of benefit to improve the accuracy of multi-dimensional supernova simulations.

\end{abstract}

\maketitle
Electron capture (EC) reactions on nuclei play important roles in the evolution of a variety of astrophysical phenomena \cite{Langanke2021}, including intermediate-mass stars~\cite{Zha2019,Doherty2017}, core-collapse supernovae (CCSNe)~\cite{Fuller1982,Janke2007,Sullivan_2016,Pascal2020}, thermal processes in neutron-star crust~\cite{Schatz2014,Chamel2021}, and nucleosynthesis in thermonuclear supernovae~\cite{Bravo2019,Iwamoto1999}. Although the results from this Letter are beneficial for a variety of astrophysical simulations, it is primarily focused on the collapse phase of CCSNe, during which ECs are responsible for regulating the electron fraction ($Y_e$) and the dynamics of the collapse. At the onset of this phase, the stellar temperature $T$, density $\rho$, and entropy $s$ are about 10 GK, 10$^{10}$ g/cm$^{3}$, and 1 $k_B$, respectively, and a nuclear statistical equilibrium (NSE) exists in the core~\cite{Bethe1990}. ECs on nuclei reduce $Y_e$, and emitted electron neutrinos carry away energy and entropy, accelerating the collapse of the core \cite{LANGANKE2000481}. It has previously been shown ~\cite{Sullivan_2016,Titus2018, Pascal2020} that the nuclei that contribute most strongly to the change in $Y_{e}$ are situated along the $N=50$ and $N=82$ shell closures near $^{78}$Ni and $^{128}$Pd, respectively. At $\rho\gtrsim$10$^{12}$ g/cm$^{3}$, the dynamical timescale of the collapse becomes shorter than the electron-neutrino diffusion timescale and the electron neutrinos become trapped~\cite{Shapiro2008,Bethe1990}. The collapse proceeds up to $\rho\gtrsim$ n$_{sat}$ $\approx$ $2.81 \cdot 10^{14}$ g/cm$^{3}$. An outward propagating shock wave is emitted at a radius where the velocity of the in-falling material equals the speed of sound. The mass of the remaining inner core is proportional to final ${Y_e}^2$~\cite{Shapiro2008,Bethe1990}.

Since the densities, and thus Fermi energies in the collapsing star, are high, ECs can occur to highly excited states in the daughter nuclei. As the temperatures are also high, excited states in the parent are thermally populated, and ECs can occur from these excited states~\cite{Bethe1979}. The EC is induced by allowed (Gamow-Teller) and forbidden transitions ~\cite{PhysRevC.101.025805,Litvinova2021,PhysRevC.102.065804}. The relevant conditions cannot be reproduced in terrestrial laboratories, and theoretical modeling is required. The theoretical models must be developed based on and tested against experimental data at zero temperature. In the past decades, experimental data from charge-exchange reactions at intermediate energies, complemented with limited information from EC/$\beta^+$-decay data, have been used for this purpose~\cite{Langanke2021}.   

\begin{figure*}[ht!]
    \centering
    \includegraphics[width=\linewidth]{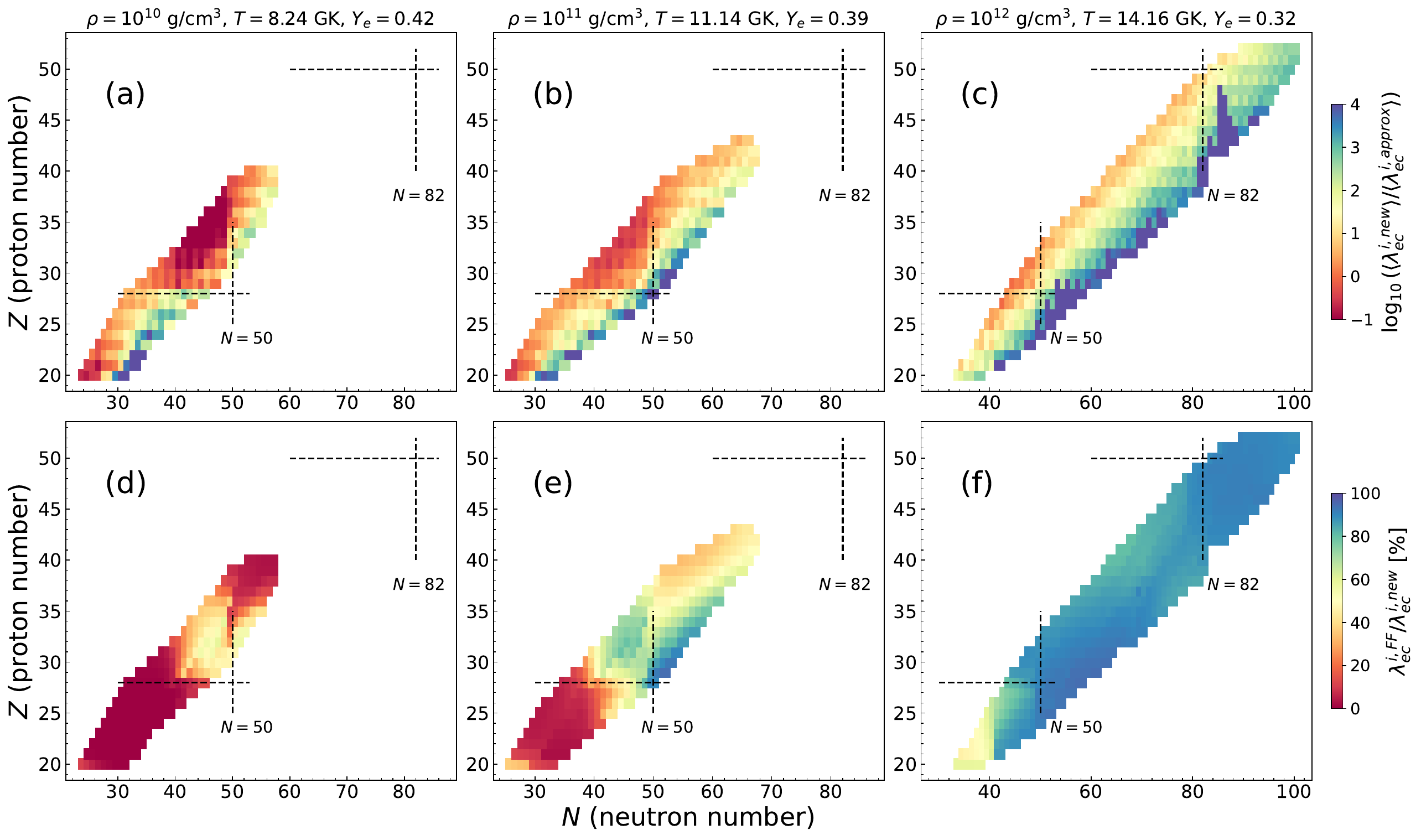}
    \caption{(a)--(c) The ratio between the NSE-averaged EDF EC rates calculated in this work with the FT-pnRQRPA  $\langle \lambda_{ec}^{i, new} \rangle$ and the rates obtained using analytic approximation from Ref. \cite{PhysRevC.95.025805}, $\langle \lambda_{ec}^{i, approx.} \rangle$. (d)--(f) Contribution of first-forbidden (FF) transitions $\lambda_{ec}^{i, FF}$ to the total EC rate, all calculated with the FT-pnRQRPA. Panels correspond to three points $(\rho, Y_e, T)$ along the CCSNe trajectory, simulated using the new EC rate set from this work. Only the nuclei with NSE abundance $Y_i > 10^{-8}$ are shown.}
    \label{fig:fig1}
\end{figure*}

The combined theoretical and experimental efforts led to the development of EC rate libraries, which employ a temperature-density grid first used by Fuller, Fowler, and Newman~\cite{FFN1980}. Subsequently, the importance of first-forbidden transitions was recognized in Ref. \cite{Cooperstein1984a}. In Ref.~\cite{JUODAGALVIS2010454}, the EC rates have been calculated for a pool of 250 nuclei using a hybrid
model of shell-model Monte Carlo (SMMC) and random phase
approximation (RPA) with the occupation numbers determined from the SMMC~\cite{PhysRevLett.90.241102} or from the Fermi-Dirac parametrization depending on the 
computational restrictions of the SMMC and sensitivity of the EC rates on detailed structure of the Gamow-Teller transitions. A pool of EC rates for more than 2200
(medium-) heavy nuclei relevant for the collapse phase, including first-forbidden transitions, was calculated using the RPA with a Fermi-Dirac parameterization for fractional occupation numbers ~\cite{JUODAGALVIS2010454}, which also contains shell-model rates from Refs. \cite{LANGANKE2000481,LANGANKE20011}.
More recently, an EC rate library~\cite{Sullivan_2016, PhysRevC.100.045805, PhysRevC.105.055801} has been developed that uses various sets of EC rates based on microscopic nuclear structure calculations that were benchmarked by experimental data complemented with a single-state approximation for nuclei~\cite{PhysRevLett.90.241102,PhysRevC.95.025805} for which microscopic calculations were not available. Benefiting from the increased computational power now available, in this work, for the first time, EC rate calculations including allowed and forbidden transitions and the effects of finite temperature are calculated within a self-consistent finite-temperature covariant energy density functional (EDF) theory and the quasi-particle random-phase approximation (QRPA) for stable and unstable nuclei with atomic numbers $Z=20$ to $Z=52$. This makes it possible to evaluate the role of ECs in the collapse phase of CCSNe within a consistent and complete nuclear physics framework. By adding the new rates to the existing library \cite{library}, the impacts of using these new EC rates are evaluated in 1D spherically symmetric simulations of the collapse phase. Multi-dimensional simulations that do not assume spherical symmetry and that are computationally demanding are necessary to fully understand CCSNe~\cite{10.1093/ptep/pts009, JAN16, MUL20, RevModPhys.85.245}. The detailed treatment of EC rates for all nuclear species in these models is not yet possible, and approximations are necessary. The library of Ref.~\cite{JUODAGALVIS2010454} was previously used to develop realistic inputs. The results from the present work enable the development of inputs for these multi-dimensional simulations based on modern microscopic theory.

The theoretical framework employed for calculating EC rates in this work is based on the relativistic EDF theory with the momentum-dependent D3C$^*$ interaction \cite{PhysRevC.105.055801,PhysRevC.71.064301,PhysRevC.93.025805}. The initial nuclear basis is determined within the finite-temperature Hartree-Bardeen-Cooper-Schrieffer (FT-HBCS) theory assuming spherical symmetry \cite{PhysRevC.102.065804,PhysRevC.96.024303}. Besides keeping the computational time manageable, such an approximation is reasonable since beyond $T \approx 2$ MeV, as in the collapse phase, most nuclei become spherical \cite{PhysRevC.109.014318}. The computational framework used in this work is detailed in Refs. \cite{Ravlic2025a,PhysRevC.105.055801}, with a key difference that the particle-particle channel employs the pairing part of the more sophisticated Gogny D1S interaction \cite{VRETENAR2005101} with overall interaction strength $V_{pair}$ determined to reproduce pairing gaps of semi-magic isotopic chains. The final states are determined within the finite-temperature relativistic QRPA in the charge-exchange channel (FT-pnRQRPA) \cite{PhysRevC.102.065804,PhysRevC.101.044305,PhysRevC.104.054318}. In order to obtain the expressions for the EC rates, we employ the current-current form of the weak-interaction Hamiltonian as in Refs. \cite{PhysRevC.6.719,PhysRevC.105.055801}. The rates include finite-momentum transfer corrections, as detailed in Ref. \cite{PhysRevC.6.719}, and we neglect effects of the electron screening on the rates, as it can be included phenomenologically \cite{JUODAGALVIS2010454}. Apart from the allowed Gamow-Teller (GT) transitions, we also consider the first-forbidden (FF) $J^\pi = 0^-, 1^-$ and $2^-$ transitions, which largely determine the EC rate for neutron-rich nuclei. We assume that nuclei are fully ionized with the Fermi-Dirac distribution of electrons in the plasma, meaning that together with the nuclear properties, the rates are a function of the temperature $T$ and a product of stellar density with the electron-to-baryon ratio $\rho Y_e$. The EC rates can be expressed through the electron neutrino spectral functions $n(E_\nu)$ \cite{PhysRevC.64.055801}
\begin{equation}
    \lambda_{ec} = \sum \limits_{J^\pi}\int \limits_0^\infty n(E_\nu, J^\pi) d E_{\nu},
\end{equation}
where $E_\nu$ is the electron neutrino energy (neglecting a small rest mass). The neutrino spectrum can be subsequently utilized to calculate the neutrino energy loss rate, as well as the average neutrino energy required for CCSNe simulations \cite{Sullivan_2016,JUODAGALVIS2010454}. Model calculations include a large set of EC rates calculated from the proton to the neutron drip-line, starting from calcium $(Z = 20)$ and terminating at tellurium $(Z = 52)$, a total of 1652 individual nuclei. The neutron(proton) drip lines are determined from the neutron(proton) chemical potential condition $\mu_{n(p)} > 0$, based on the FT-HBCS calculation with the D3C${}^*$ interaction. The odd-$A$ and odd-odd systems are constrained with the particle number, similar to Refs. \cite{PhysRevC.105.055801,PhysRevC.93.025805}. The considered temperatures span a range from 0.01 to 50 GK, with a respective range in densities from $\rho Y_e = 10^1$ to $10^{14}$ g/cm${}^3$, enough to consider the CCSNe trajectories where electron captures are of relevance.

\begin{figure}
    \centering
    \includegraphics[width=0.85\linewidth]{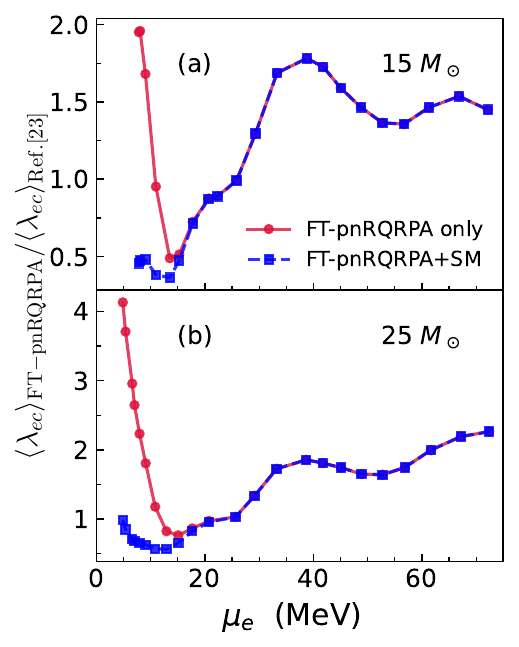}
    \caption{The ratio between the total NSE-averaged rates calculated in this work, $\langle \lambda_{ec} \rangle_{\rm FT-pnRQRPA}$, with unscreened rates from Ref. \cite{JUODAGALVIS2010454}, $\langle \lambda_{ec} \rangle_{\rm Ref.[23]}$, along two CCSNe trajectories presented in Ref. \cite{JUODAGALVIS2010454}, corresponding to either $15M_\odot$ (a) or 25$M_\odot$ (b) progenitor, and parameterized by the electron chemical potential $\mu_e$. We compare the results from this work by either using only the FT-pnRQRPA EC rates (solid line) or by supplementing the FT-pnRQRPA rates with the shell-model calculations, where available (dashed line). }
    \label{fig:fig2}
\end{figure}

\begin{figure*}[t!]
    \centering
    \includegraphics[width=\linewidth]{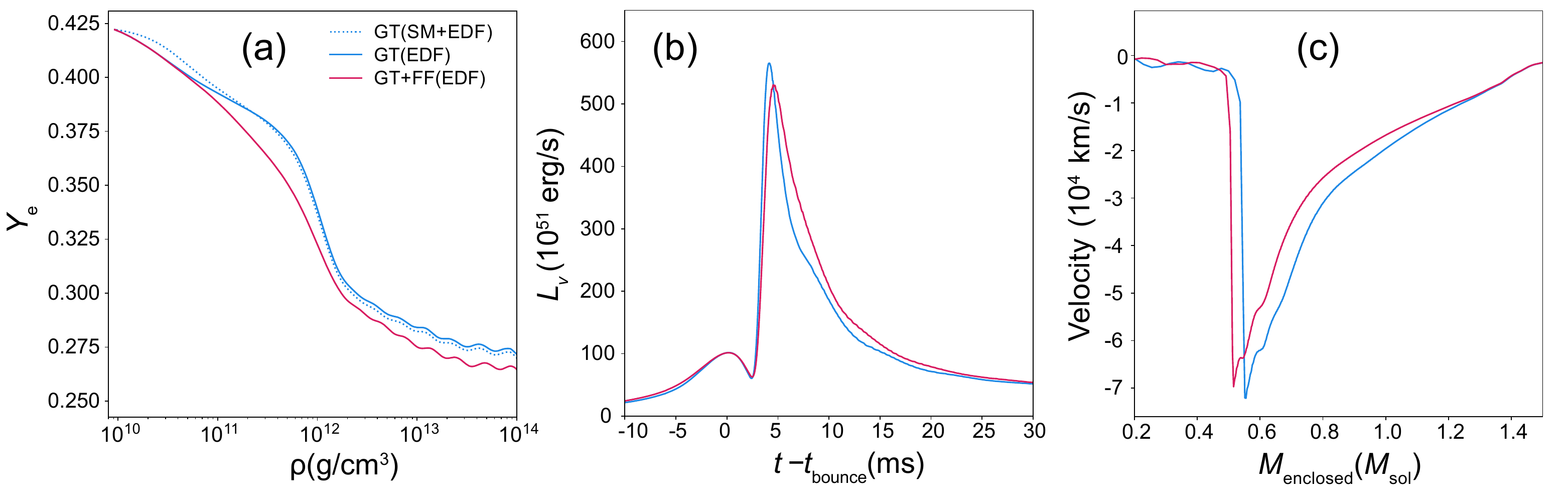} 
    \caption{The results for the main CCSNe observables obtained for 3 different simulations with GR1D and NuLib using s15WW95 progenitor with SHFo EOS. Results are shown for (a) the density evolution of electron-to-baryon ratio $Y_e$, (b) the peak neutrino luminosity $L_\nu$ at 500km from the core as a function of time, and (c) the central velocity as a function of the enclosed mass. Simulations are performed with EC rates calculated using the FT-pnRQRPA, including GT [GT(EDF)], and GT+FF transitions [GT+FF(EDF)] (solid lines). In addition, in panel (a) one simulation, [GT(SM+EDF)], employs the dataset from the present work including GT transitions, but using shell-model rates where available (dotted line). }
    \label{fig:fig3}
\end{figure*}

The nuclear composition in the presupernova matter is obtained by considering the NSE \cite{PhysRevC.82.065801}. Therefore, it is most relevant to check the rates of nuclei that have non-negligible NSE abundance. To get the NSE distribution, we employ the framework of Refs. \cite{Seitenzahl2008,JUODAGALVIS2010454}, with the Timmes and Arnett equation of state (EOS) from Ref. \cite{Timmes1999}. The NSE is determined by three parameters: stellar density $\rho$, electron-to-baryon ratio $Y_e$, and temperature $T$, along a CCSNe trajectory. Having the NSE abundances $Y_i$ for each nuclear species (which are normalized as $\sum \limits_i Y_i = 1$), we can define the NSE-averaged rates as
\begin{equation}\label{eq:averaging}
    \langle \lambda_{ec}^i \rangle = Y_i \lambda_{ec}^i.
\end{equation}
Due to its simplicity, many CCSNe simulations employ analytic parameterizations for EC rates, assuming that rates depend strongly on EC $Q$-values through a series of Fermi integrals and fit coefficients to shell-model calculations in the $pf$-region \cite{PhysRevLett.90.241102, PhysRevC.95.025805}. Such an approach omits detailed nuclear structure effects and has limited accuracy for neutron-rich nuclei where shell-model calculations are unavailable and extrapolations are necessary. Here, we compare our rates, $\langle \lambda_{ec}^{i, new} \rangle$, with the approximation from Ref. \cite{PhysRevC.95.025805}, $\langle \lambda_{ec}^{i, approx} \rangle$, averaged over the NSE distribution. The ratio between our NSE averaged rates and the approximation is shown in Fig. \ref{fig:fig1}(a)--(c) for three points along the CCSNe trajectory. Starting from the FT-pnRQRPA rate table \cite{library}, the rates are interpolated using monotonic splines as discussed in Ref. \cite{Sullivan_2016}. We observe that with increasing stellar density $\rho$, the distribution of nuclei shifts to higher mass. At $\rho = 10^{10}$ g/cm$^3$, the NSE composition is dominated by nuclei up to $Z \approx 40$, and our calculations indicate higher rates than the approximation for more neutron-rich nuclei. At $\rho = 10^{11}$ g/cm${}^3$, the ratio between the two rates for most nuclei is within one order of magnitude, with larger differences observed for more neutron-rich nuclei. Finally, at $\rho = 10^{12}$ g/cm${}^3$, the NSE composition reaches $N = 82$ region, and considerable differences between the FT-pnRQRPA rates and the approximation are obtained; for most of nuclei the new rates are larger, even up to 4 orders of magnitude for neutron-rich nuclei. The large discrepancy between the FT-pnRQRPA calculations and the approximation showcases a necessity to consider the rates within a microscopic framework, incorporating crucial effects of temperature on nuclear excitations. Another possible reason for the discrepancy is the significant contribution of FF transitions. In Fig. \ref{fig:fig1}(d)--(f), we display the percentage of FF transitions in the total rate. At $\rho = 10^{10}$ g/cm${}^3$, the EC rates are mostly dominated by allowed transitions. However, as the density increases to $10^{12}$ g/cm${}^3$, close to the neutrino trapping, the FF transitions dominate the rates for the whole NSE composition. In the following, we compare our rates with those from Ref. \cite{JUODAGALVIS2010454}, used in numerous forefront CCSNe simulations. Since the EC rates in Ref. \cite{JUODAGALVIS2010454} are presented in terms of the NSE-averaged total rate, we perform the NSE-averaging as in Eq. (\ref{eq:averaging}) and sum over all individual nuclei. The comparison of unscreened averaged EC rates is shown in Fig. \ref{fig:fig2}(a)--(b), for two CCSNe trajectories outlined in Ref. \cite{JUODAGALVIS2010454} and parameterized in terms of the electron chemical potential $\mu_e$. To better visualize the differences, we take the ratio between the respective NSE-averaged rates. The rates in Ref. \cite{JUODAGALVIS2010454} employ the quenching of GT strength of around $ 0.5$, while in this work we apply the quenching of around $ 0.64$, which should account for around 20\% difference between the rates. Sizable differences can be observed between the two rate sets, in particular up to $\mu_e \approx 10$ MeV, roughly corresponding to temperatures up to $T \approx 10$ GK and $\log_{10} \rho Y_e \approx 9$. The differences mainly originate from the fact that our rates are larger than the shell-model rates under these conditions for $pf$-shell nuclei. To better demonstrate this, we consider the comparison by using our FT-pnRQRPA rates and supplementing them with the shell-model rates where available. With this modification the combined FT-pnRQRPA+SM rates underestimate the rates from Ref. \cite{JUODAGALVIS2010454} up to $\mu_e \approx 25$ MeV for both progenitors. At higher temperatures and densities the NSE pool of nuclei goes beyond the $pf$-shell, and supplementing rates with shell-model data play no effect. At these conditions, the FT-pnRQRPA rates are mostly larger than the rates from Ref. \cite{JUODAGALVIS2010454} by up to a factor of 2.

The CCSNe simulations are performed using the GR1D code, as in Refs. \cite{Sullivan_2016,PhysRevC.105.055801,PhysRevC.100.045805}. The GR1D considers early and post-bounce stages in spherical symmetry with general relativistic hydrodynamics, while the neutrino transport is handled through the NuLib library \cite{O_Connor_2015}. All simulations consider a 15-solar-mass, solar-metalicity, star progenitor (s15WW95 \cite{1995ApJS..101..181W}) with SHFo EOS to determine the NSE \cite{Steiner_2013}. We performed CCSNe simulations with the new FT-pnRQRPA EC rates for 1652 nuclei based on relativistic EDF theory, either including only the GT contribution [GT(EDF)] or GT with FF transitions [GT+FF(EDF)]. As a benchmark, we performed an additional simulation, but employing shell-model rates \cite{LANGANKE2000481,SUZUKI2022103974,Suzuki_2016,ODA1994231} where available ($sd$- and $pf$-shell nuclei), and otherwise using the rates calculated in this work [GT(SM+EDF)]. Where available, the shell-model calculations generally achieve the highest accuracy \cite{Cole12} for nuclei up to the $pf$-shell but are not available for heavier nuclei and away from the valley of stability, and do not include contributions from the forbidden transitions. As noted in Ref. \cite{Dzhioev2010a}, QRPA calculations cannot capture additional correlations that are included in shell-model calculations, which are important for overcoming cross-shell gaps \cite{ZHI2011172}. Therefore, they tend to underestimate the shell-model rates at lower temperatures and densities. However, at higher temperatures and densities, relevant for CCSNe evolution --- $T \approx 10$ GK and $ \rho Y_e \approx 10^{10}$ g/cm${}^3$, where the rates depend on the bulk of the underlying strength function, both become comparable. At these specific conditions, the EC rates based on FT-pnRQRPA, presented in this work, are somewhat higher than the presently published shell-model rates. Detailed comparison can be found in Refs. \cite{Ravlic2025a,PhysRevC.105.055801,PhysRevC.102.065804}.

First, in Fig. \ref{fig:fig3}(a), we show the density evolution of the electron-to-baryon ratio $Y_e$. Up to $\rho = 10^{11}$ g/cm${}^3$, we observe almost no impact of FF rates on the $Y_e$, as the pool of relevant nuclei consists of light and medium-mass nuclei for which mostly GT transitions are relevant. However, as the density increases beyond $10^{11}$ g/cm${}^3$ there is a significant impact of FF transitions, driving additional deleptonization and decreasing $Y_e$. As can be seen in Fig. \ref{fig:fig1}(e), at $\rho = 10^{11}$ g/cm${}^3$, the NSE averaged pool of nuclei extends up to $Z = 45$, with ECs on nuclei beyond $Z = 30$ having a significant contribution from the FF transitions. Ultimately, including FF transitions results in a 3\% reduction of $Y_e$ close to the saturation density. By including the shell-model rates [GT(SM+EDF)], small differences in $Y_e$ can be noticed up to $10^{11}$ g/cm${}^3$. As explained above, under these conditions they result in a slightly higher $Y_e$ because the GT(EDF) rates tend to be larger than the shell-model rates, increasing the deleptonization. However, for $\rho > 10^{11}$ g/cm${}^3$, the trajectory is almost independent of whether we use the SM rates or EDF rates for $pf$-shell nuclei. The density evolution of $Y_e$ directly affects the time evolution of the neutrino luminosity $L_\nu$, shown in Fig. \ref{fig:fig3}(b). Before the bounce at $t \approx t_{bounce}$, the EC is dominated by $pf$-shell nuclei, and we observe almost no influence of FF transitions. However, due to faster deleptonization with FF transitions included, the peak neutrino luminosity is delayed by around 0.7 ms and decreases in amplitude by almost 5\%, resulting in an overall considerably different luminosity profile. Finally, in Fig. \ref{fig:fig3}(c), we display the central velocity as a function of the enclosed mass, which determines the effective mass of the underlying proto-neutron star (PNS). Including FF transitions within the new EC dataset leads to around a 7\% reduction in the mass of the PNS.

\begin{figure}
    \centering
    \includegraphics[width=\linewidth]{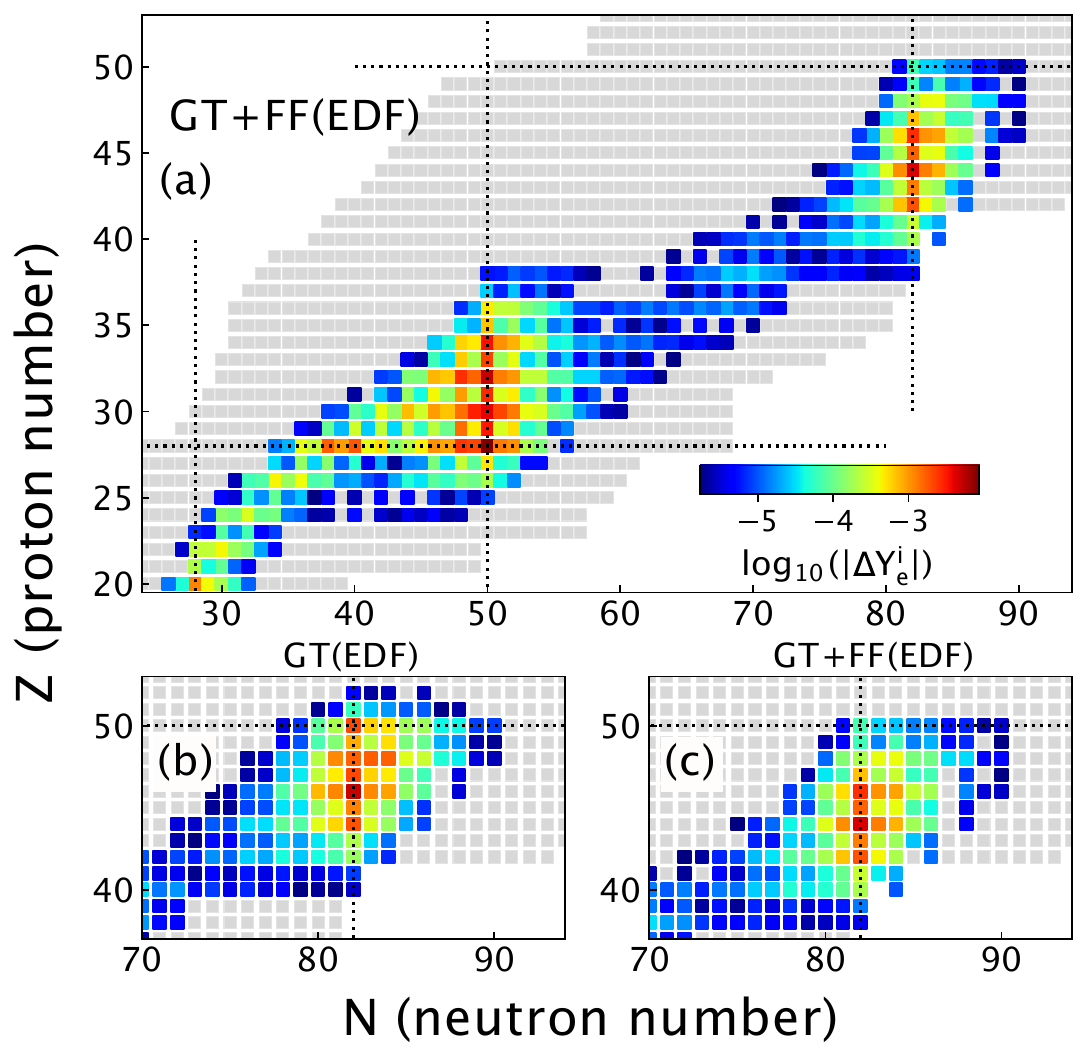}
    \caption{The electron-to-baryon ratio change $|\Delta Y_e^i|$ for individual nuclei, integrated from the onset of the collapse up to the time of neutrino trapping ($\rho \approx 10^{12}$ g/cm${}^3$). Nuclei with higher $|\Delta Y_e^i|$ have a higher influence on driving the deleptonization and larger impact on CCSNe dynamics. (a) The results showing full FT-pnRQRPA EC rates including both GT and FF transitions. Lower panels display $|\Delta Y_e^i|$ for nuclei near $N = 82$ shell closure without (b) and with (c) FF transitions.}
    \label{fig:fig4}
\end{figure}

Finally, we aim to quantify the impact of individual nuclei on the CCSNe dynamics. The study in Ref. \cite{Sullivan_2016} determined two regions of the nuclear chart near $N = 50$ and $N = 82$, which have the highest contribution to the deleptonization process $\Delta Y_e^i$. Unfortunately, due to a lack of theoretical calculations in these regions, the study utilized analytic approximations for the rates. Therefore, we implement the new set of FT-pnRQRPA EC rates in the analysis of the impact of individual nuclei on the deleptonization $\Delta Y_e^i$, calculated by summing $|\dot{Y}_e|$ for each individual nuclear species and integrating it from the onset of collapse phase up to the neutrino trapping $t = t_{trapping}$. Results are shown in Fig. \ref{fig:fig4}(a) for the EC rate set calculated with the FT-pnRQRPA, including GT and FF transitions. Simulations employing the new rate set confirm the results obtained in Ref. \cite{Sullivan_2016}, that regions around $N = 50$ and $N = 82$ have the most significant impact driving the deleptonization. Including FF transitions plays almost no role for nuclei around $N = 50$ shell closure. However, for the $N \approx 82$ nuclei, we notice a shift towards lower-$Z$ nuclei as indicated in Fig. \ref{fig:fig4}(c) which contains FF contribution, compared to Fig. \ref{fig:fig4}(b) with only GT. In particular, the deleptonization in ${}^{132}$Sn changes by an order of magnitude once the FF transitions are included. We would like to highlight that present results are based on employing the nuclear composition from Ref. \cite{Hempel2010}. While the prescription from Ref. \cite{Hempel2010} is widely-used in determining nuclear composition, results for most relevant nuclei driving the deleptonization could be modified by improvements to nuclear inputs and partition function treatments.

In conclusion, we have developed a set of stellar density- and temperature-dependent EC rates for nuclei ranging from $Z=20$--52 within the self-consistent finite-temperature covariant EDF theory based on the FT-pnRQRPA, which includes contributions from allowed and first-forbidden transitions from ground and thermally-populated excited states. Presented calculations provide a fundamental improvement over previous EC rate estimates by not using approximations and extrapolations that do not consider detailed nuclear structure effects. The implementation of the new rates in astrophysical simulations of CCSNe, compared to the previous version of weak-rate library \cite{library}, indicates that the EC rate is enhanced, resulting in a reduction of the electron fraction at saturation, the peak neutrino luminosity, and the enclosed mass at the bounce. By employing a consistent microscopic description of the EC rates, we have demonstrated that nuclei most critical to the deleptonization process near $N = 82$ shell closure become less proton rich, which could provide guidance for future experimental efforts.
The EC rates established in this work provide an essential foundation for advancing the accuracy of the future multi-dimensional supernova simulations, and better understanding the evolution of stellar explosions.

\textbf{\textit{Data availability}} The data that support the findings of this article are
openly available \cite{library}.

\textbf{\textit{Acknowledgement.}} We thank Gabriel Mart\'inez-Pinedo for insightful discussions and providing the NSE code. We also thank Toshio Suzuki for his insights and support.
This work was supported by the U.S. Department of Energy under Award No. DOE-DE-NA0004074 (NNSA, the Stewardship Science Academic Alliances program), the U.S. National Science Foundation under award number NSF-2209429, Windows on the Universe: Nuclear Astrophysics at FRIB, and
by the Croatian Science Foundation under the project number HRZZ-IP-2022-10-7773. This work was supported in part through computational resources and services provided by the Institute for Cyber-Enabled Research at Michigan State University. We also acknowledge support by the U.S. National Science Foundation under Grant No. PHY-1927130 (AccelNet-WOU: International Research Network for Nuclear Astrophysics [IReNA]).

%

\end{document}